# Wave-front controlled attosecond time domain interferometry

Zhen Yang[1], Wei Cao[1, *], Yunlong Mo[1], Huiyao Xu[1], Kang Mi[1], Pengfei Lan[1], Qingbin Zhang[1, *] and Peixiang Lu[1,2, *]

[1]*Wuhan National Laboratory for Optoelectronics and School of Physics, Huazhong University of Science and Technology, Wuhan, 430074, China*

[2]*Hubei Key Laboratory of Optical Information and Pattern Recognition, Wuhan Institute of Technology, Wuhan 430205, China*

*email: weicao@hust.edu.cn; zhangqingbin@hust.edu.cn; lupeixiang@hust.edu.cn

**Interferometry, as the key technique in modern precision measurements, has been used for length diagnosis in the fields of engineering metrology and astronomy. Analogous interferometric technique for time domain precision measurement is a significant complement to the spatial domain applications and requires the manipulation of the interference on the extreme time and energy scale. Here we present an all-optical interferometry for high precision measurement with attosecond temporal and hundreds of meV energy resolution. The interferometer is based on laser driven high order harmonics which provide a robust sequence of attosecond temporal slits. As applications, we reconstruct the waveform of an arbitrarily polarized optical pulse using the time resolving capability, and capture the abnormal character of the transition dipole near a Cooper minimum in argon using the energy resolution provided. This novel attosecond interferometry opens the possibility for high precision measurement in time energy domain using an all-optical approach.**

Young's double slit or the equivalent two arm interference experiment, as a direct proof of the wave nature of light or matter, has been used to test the basic principles in quantum mechanics [1-4]. Apart from the pivotal role it played in understanding the quantum world, such interferometric scheme is also a sensitive tool for applications in high precision measurement. For example, photoionization from molecules is equivalent to an atomic scale interferometer [5-7]. The momentum distribution of photoelectron in the molecular frame, as Fourier transformed to the real space, offers superb spatial resolution for accessing chemical bond length [6,7]. A more versatile approach for interferometer-based precision measurement is to change the phase of individual arms by inserting objectives to be investigated. In this way the wave-front of the incoming waves at the slits are effectively manipulated, leading to profound change in the final interference pattern, from which quantities related to the inserting objectives can be accurately extracted. Pioneering experiments have successfully measured the Newton's constant G [8], fine structure constant [9] and index of refraction of dilute gases [10] using atomic interferometers.

Following the same spirit of wave-front manipulation scheme, efforts has also been put into extending precision measurement to time domain for exploring exceedingly fast processes [11-13]. Recently an all-optical interferometer combining isolated attosecond pulse and the Young's configuration was developed. Two extremely ultraviolet (EUV) isolated attosecond pulses separated spatially interfere in the far field. By introducing a weak optical pulse to perturb one interference arm, the angular distribution of the far field pattern acquired a displacement which was used to mapped out the waveform of a vectorial optical field [13]. However, the high temporal resolution in this measurement is achieved at the trade-off of spectral resolution, preventing energy resolved dynamic studies that are essential for enhancing our understanding of light-matter interaction. The contradiction between the temporal and spectral resolution may be circumvented with a pulse train. As demonstrated by M Isinger *et al.* [14], laser assisted photoemission of neon using an attosecond pulse train creates photoelectron spectrogram bearing high temporal and adequate spectral resolution simultaneously. Different photoionization pathways that are nearby in energy are disentangled, allowing unambiguous determination of the photoemission time delay, thereby solving a long standing puzzle [15-17]. Therefore, all-optical interferometer can be greatly beneficial from the use of pulse trains and holds the promise of extending high precision measurement to the time-energy domain. In this work, we present an all-optical time domain few-slit interferometer with an attosecond pulse train. By manipulating the temporal wave-front of the pulse train in a controlled manner we demonstrate its applications in high precision measurement in both time and energy domain. In the first application, we successfully recovered the waveform of the perturbing field and implemented an all-optical petahertz oscilloscope. In the second application, the intrinsic energy resolution provided by the interferometer allows one to perform energy-resolved analysis of the involved dynamics. The reshaping of attosecond pulses due to the presence of a Cooper minimum in the generating medium is captured.

**Principle of wave-front controlled attosecond few slit interferometry**

The time domain interferometer is based on the laser driven EUV radiation [18,19]. The principle of the method is illustrated in figure 1. When phase matching condition is fulfilled EUV radiation is then launched within a few attosecond time windows near the peak of the driving field [18,20]. Under the pulse train configuration, the interferometer is analogous to the conventional spatial double-slit experiment except that the temporal slits are used for diffraction instead. This conceptual experiment has been successfully demonstrated in matter waves using nanosecond [21], femtosecond [22] and attosecond time slits [23]. As a result of interference between radiation from different time windows, localized frequency components are created and are known as the high order harmonics [24,25]. When a weak signal pulse is applied, it perturbs the harmonic generation process and introduces a phase variation in the wave-front of the time slits [26]. As a result, each high order harmonic experiences a prominent energy shift upon which temporal structure of both the perturbing signal and the attosecond slits is imprinted.

The harmonic energy shift can be derived within the framework of strong field approximation [28]. High order harmonics are emitted every half optical cycle of the driving field. For simplicity, we consider a double slit interferometer configuration. The phase difference of the harmonic radiating from two consecutive attosecond slits in absence of the perturbing signal is $\Delta\phi(\omega) = \frac{\omega T}{2} + \pi - \delta$, where T is the optical cycle of the driver and $\delta$ accounts for the non-adiabatic effect of the ultrashort pulse [27]. When a signal pulse is present, it introduces an additional phase on each attosecond slit and the total phase difference is (see section I the supplementary):

$$\Delta\Phi(\omega,\tau) = \frac{1}{72}\alpha E_0 \omega_0 t_d^4 E_s(\tau + \Delta) + \frac{1}{72} E_0 \omega_0 t_d^4 E_s(\tau) + \Delta\phi(\omega)$$

Where $E_0$ and $E_s$ are the amplitudes of the electric field of the driver and signal, respectively. $t_d$ stands for the excursion time in the three step model of high harmonic generation. $\tau$ is the delay between the signal and driving pulse, $\Delta$ is the time separation between the consecutive attosecond pulses. The coefficient $\alpha$ is introduced to consider the intensity variation of the driving field within an optical cycle and is close to 1 in general. The EUV spectral maxima are then determined by the constructive interference condition $\Delta\Phi(\omega,\tau) = 2m\pi$ with m an integer, and the harmonics will peak at:

$$\omega = (2m-1)\omega_0 + \frac{2\delta}{T} - \frac{E_0\omega_0 t_d^4}{36}\big(\alpha E_s(\tau + \Delta) + E_s(\tau)\big) \qquad (1)$$

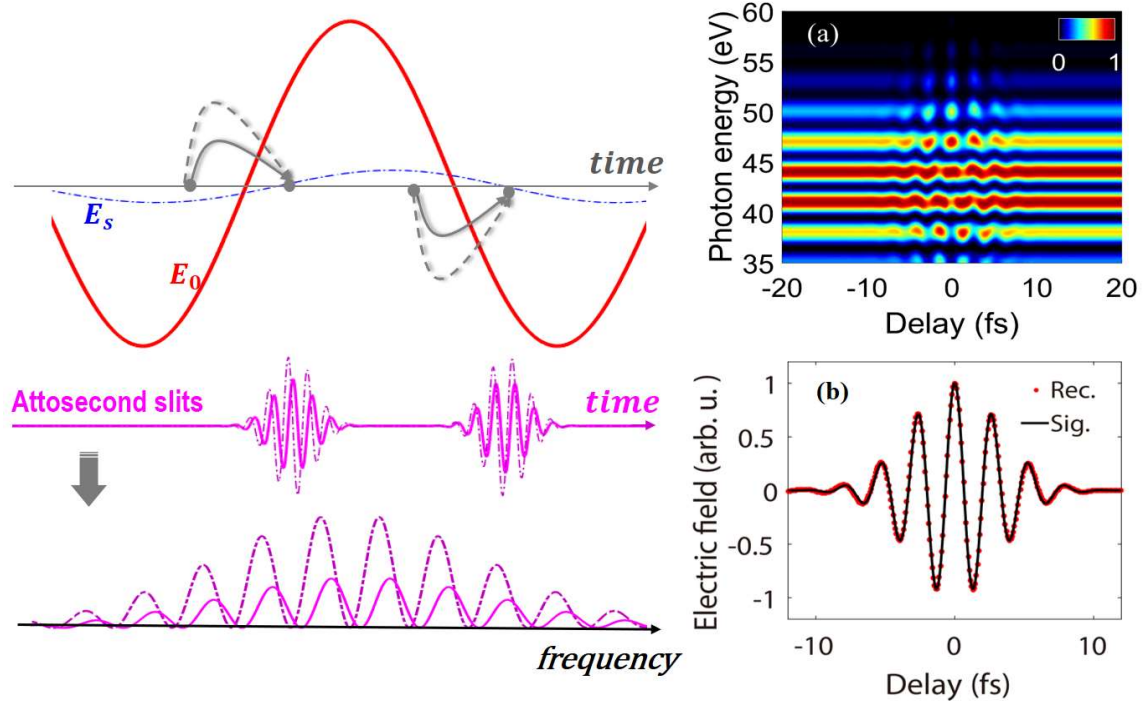

**Figure 1**. **Principle of wave-front controlled attosecond few slit interferometry**. Strong driving pulse $E_0$ generates high order harmonics every half optical cycle of the driver, the attosecond bursts (pink solid lines) maintains mutual coherence and is equivalent to a Young's interferometer with the attoecond pulses as the slits. The interference leads to fringes in the frequency domain. When a weak signal pulse $E_s$ is synchronized with the driver, it perturbs the electron trajectories (grey curved arrows) for harmonic generation and imposes additional phase in each attosecond slit, this will induce a shift of the interference pattern in the frequency domain. (a)The simulated harmonic spectrum using strong field approximation indicates that the energy shift of the harmonics is sensitive to the relative delay between the driving field and signal pulse. The delay dependent energy shift of a single harmonic can be expressed as $\sigma(\tau) \propto E_s(\tau) + \alpha E_s(\tau + \Delta)$ (see the text), which can be used to directly reconstruct the electric field of the signal pulse. (b) The reconstruction of waveform of the signal using harmonic around 42eV, the reconstructed (red dotted lines) and original (black solid lines) field agree with each other.

The first term in equation (1) indicates that odd harmonics are produced by a continuous laser. The second term corresponds to the non-adiabatic effect induced blue shift of the harmonics. The last term is a quantitative description of the influence of the perturbing signal pulse. The signal pulse induced energy shift is then:

$$\sigma(\tau) \propto E_s(\tau) + \alpha E_s(\tau + \Delta) \qquad (2)$$

Equation (2) shows that the shift of the harmonic peak is proportional to a linear combination of two delayed signal pulses and elaborates the quantitative connection between the interference pattern and the electric field of the perturbing pulse. An immediate application of this interferogram is for real time probing the electric field of an optical pulse. Figure 1(a) shows the delay dependent high harmonic spectrum calculated using the strong field approximation [28]. As the two pulses are well separated in time, the signal played a negligible impact on the harmonic spectrum since the signal pulse is too weak to contribute to harmonic yield alone. When the delay approaches the overlap region, delay dependent oscillations in both the flux and central frequency of each harmonic sets in. The profound energy shift of each harmonic in the overlap region is governed by equation (2). Following a trivial Fourier analysis, the electric field of the signal pulse can be reconstructed (section II of supplementary). The excellent agreement between the input and the reconstructed signal electric field as shown in figure 1(b) verified the feasibility of the current method.

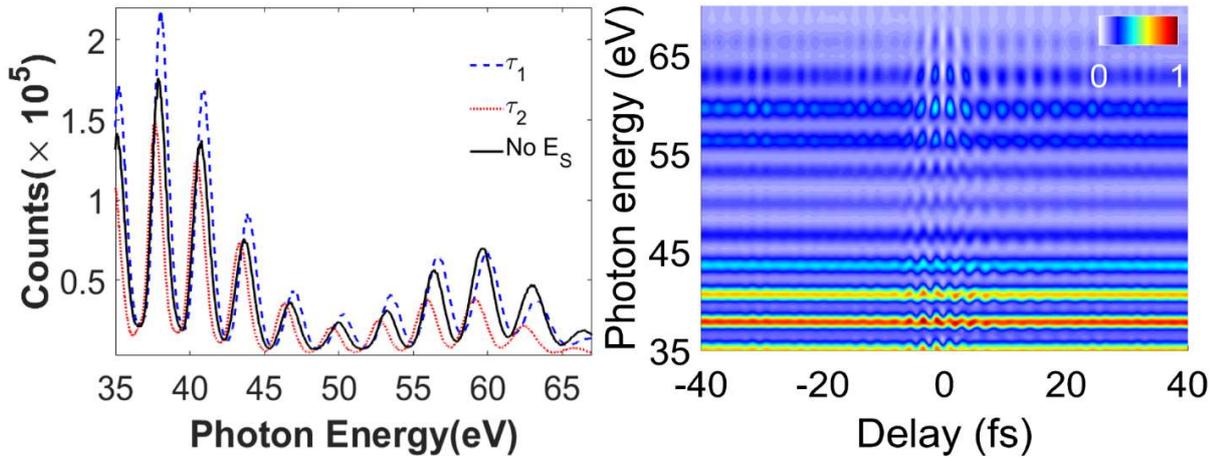

**Figure 2. Experimental confirmation of the attosecond interferometry**. (a) Measured high harmonic spectra without (black solid line) and with the perturbing signal pulse. The harmonics experience maximal blue (blue dashed line) and red (red dotted line) shift at different delays due to interference between consecutive attosecond slits. (b)Normalized two dimensional spectrogram of the high order harmonic radiation. The measured interference pattern resembles the theoretical prediction.

**Waveform sampling of an optical pulse**

In the experiment, a 7fs 760nm few-cycle laser pulse is used for high harmonic generation. The focusing geometry of the driver is carefully tuned to phase match the short trajectory such that a robust attosecond pulse train is formed. A weak signal pulse with an intensity below 1% of that of the driver is picked up to perturb the harmonic generation process. Figure 2(a) shows the impact of the weak signal field on the measured high harmonic spectrum. As the signal pulse is in (out of) phase with the driver, the spectral cutoff as well as the individual harmonic peak experience a blue (red) shift with respective to the case where no signal pulse is present. The constructive (destructive) interference between the signal and driver can increase (decrease) the peak intensity of the combined electric field and leads to the extension (reduction) of the spectral cutoff [29,30]. However, the shift of each harmonic peak is a clear signature of phase variation imposed in the attosecond pulse train by the signal field. The delay dependent high harmonic spectrum is shown in figure 2(b), it resembles the pattern that is predicted by the quantum simulation as shown in figure 1(a). Figure 3 illustrates the reconstruction of the waveform of the signal used in the experiment. Harmonic of argon around 50 eV is chosen for analysis. Figure 3(b) shows the power spectrum of the delay dependent energy shift. The spectrum spans from 1.3 to 2.3 eV and corresponds to the spectral range covered by the signal field. The minimum around $\omega_d \sim 1.6$ eV is due to destructive interference of the two delayed signals depicted in equation (2), from which the time interval of two consecutive attosecond slits $\Delta$ can be accurately calculated: $\Delta = \frac{\pi}{\omega_d}$. With $\Delta$ known, it is straightforward to extract the complete information of the electric field, which is illustrated in figure 3(c).

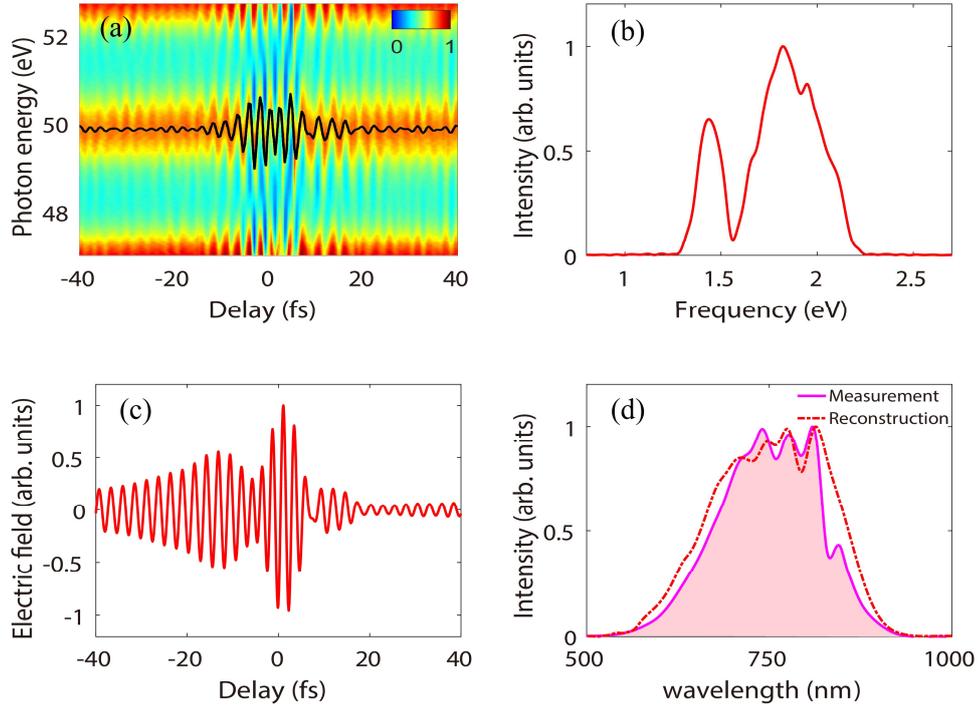

**Figure 3. The reconstruction of the waveform of a linearly polarized signal pulse**. (a) Enlarged spectrogram of the high order harmonic radiation in figure 2b. The black solid line indicates the centroid of the harmonic near 50 eV. (b) Power spectrum of the black solid line of (a). (c) Retrieved electric field of the signal pulse following the method introduced in section II of the supplementary material. (d) The reconstructed (red dashed line) and the measured (pink solid line) spectrum of the signal.

The above analysis is based on a double-slit configuration. In a real experiment a few cycle intense driver normally opens more than two attosecond slits depending on the harmonic order. As is shown in figure 2 the width of the individual harmonic becomes progressively wider as the photon energy increases, indicating that the number of attosecond pulses increases with the harmonic order decreasing. To prove the validity and generality of the reconstruction scheme, we calculated the energy shifts for different numbers of slits by applying an appropriate window function. The calculation results show that the energy shift traces are almost identical for a slit number up to five (section II of the supplementary material). It means that equation (2) still holds for a few-slit configuration. This rests on the fact that in a multi-slit experiment the dominant contribution to the fringes is the interference between two adjacent slits. Therefore, the current scheme for waveform reconstruction applies for a few-slit interferometer and can be easily implemented by using by a few-cycle driver pulse. Note that we used a carrier envelope phase(CEP) un-stabilized laser in the experiment, the random phase variation between laser shots indicates that the reconstructed envelope in figure 3 is an CEP averaged quantity. However, since the signal pulse is phase locked with the driver at high precision (RMS<15as, see section IV of supplementary material), the carrier which is the most important feature of a short pulse is fully reconstructed. Further measurements showed that material induced dispersion of the short pulse can be accurately recovered using the current method (section III of supplementary material). It is demonstrated that the waveform of a linearly polarized optical pulse can be measured. This method can also be generalized for reconstructing signals with arbitrary state of polarization as well. Since the polarization component of the signal perpendicular to the driver makes a negligible contribution to the phase variation of attosecond slits under the intensity used [13], the two orthogonally polarized components of the signal pulse can be reconstructed independently by simply rotating the polarization of the signal by 90 degrees. Figure 4 shows the reconstructed waveform of two elliptically polarized perturbing signals. It shows that our method is sensitive to the phase and amplitude ratio of the two components of the signal field, and can measure waveform of the signal field with great accuracy. The detailed experimental method can be found in section IV of the supplementary material. We have thus readily implemented a petahertz optical oscilloscope.

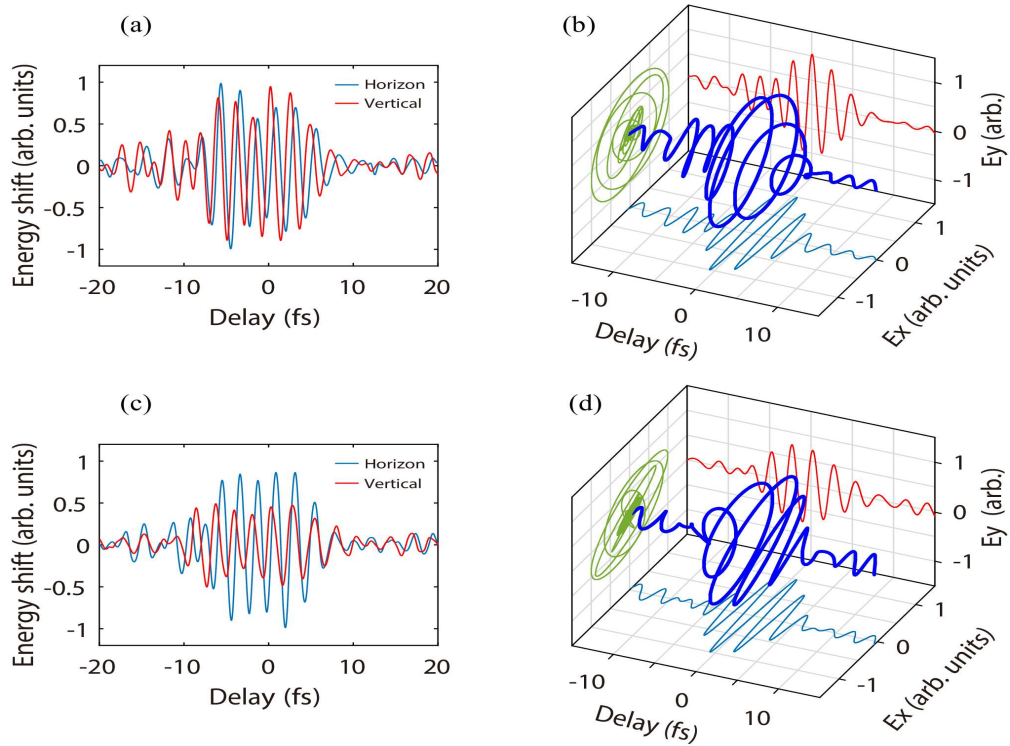

**Figure 4**. **The reconstruction of the waveform of vectorial signal fields**. The measured delay dependent energy shifts of harmonic near 50 eV with the two polarized components parallel to the polarization direction of the driving field for a circularly polarized (a) and an elliptically polarized (c) signal pulse. The reconstructed electric field for the circularly polarized signal (b) has an ellipticity close to 1. The reconstructed electric field for the elliptically polarized signal (d) has an ellipticity of 0.65 as compared to 0.58 in theory.

**Probing abnormal structure around a Cooper minimum**

Optical oscilloscopes based on high order harmonics have been reported previously using isolated attosecond pulse [13,26,31]. In the current scheme, this is realized by using an attosecond pulse train instead. The spectral resolution is crucial for performing channel resolved measurements that are generally hindered in case of isolated attosecond pulse due to the time-energy uncertainty relationship. The intrinsic energy resolution provided in the current interferometric technique therefore implies a second application for probing structural information. The separation of the two consecutive slits $\Delta$ is trackable from the Fourier transform of the energy shift trace of a single harmonic. When multiple harmonics are considered simultaneously, an energy resolved analysis of the temporal structure of the attosecond slits can be carried out. Figure 5 illustrates the spectral analysis of the delay dependent energy shifts for harmonics from 36 to 59 eV. Two different harmonic generation targets, argon and neon, have been investigated. In the case of neon (figure 4(a)), the destructive interference minimum is located at ~1.6 eV, and is almost independent on the harmonic order. This is expected for high harmonic generation from targets such as neon with regular atomic structure, i.e. the phase of the recombination dipole matrix element is smooth over the range of harmonic spectrum of interest. It reflects that the generated attosecond pulse train has a regular temporal structure where constant pulse separation maintains over a broad range of photon energy. When the generating target is argon (see figure 5(b)), an abnormal shift of the destructive interference minimum is observed centered around 50eV over a range of more than 10eV. This minimum is shifted towards a higher frequency corresponding to a reduced separation between two attosecond pulses around this photon energy. This reshaping of the attosecond pulse is likely due to the abnormal atomic structure of the generating targets. It is known that a Cooper minimum exists around 50eV in argon harmonic spectrum and is caused by the interference between different channel contributions in the recombination step [32]. The total recombination dipole undergoes a phase jump near this minimum and alters the time structure of the radiating EUV pulses [33]. To prove that it is the Cooper minimum (CM) that leads to the abnormal shift in figure 5(b), we performed a model calculation based on SFA. In the simulation we used a dipole similar to ref [33] to mimic the interference between the s and d contributions (see supplementary section V). The calculation result is consistent with the experimental results and thus provides evidence that the current interferometry is a sensitive device for structural determination.

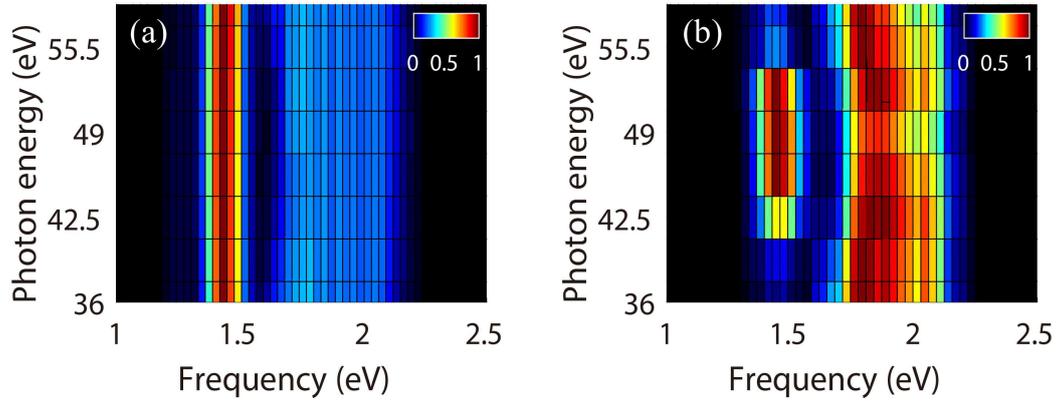

**Figure 5. Comparing two generating targets.** Fourier transform of energy shift along the delay axis for harmonics from 36 to 59 eV. Two different generating targets are used: neon(a) and argon(b). For each harmonic (vertical axis), the spectrum has been normalized to its maxima.

## Conclusion

In summary, we present an all-optical Young's interferometer using attosecond temporal slits and explored its application for high precision measurement. High order harmonics driven by an intense few-cycle optical pulse is naturally a perfect time-energy domain few-slit interferometer. By synchronizing a delayed perturbing signal, the control of the wave front of the temporal slits is realized which leads to an energy shift of individual harmonics. This controlling scheme provides a two-dimensional interferogram bearing both superb time and adequate energy resolution for precision measurements. An intuitive analytical formula is derived to depict the perturbing field induced energy shift. The delay dependent shift of a single harmonic provides sufficient time resolution for attosecond sampling of a petahertz electromagnetic field. When the shifts of multiple harmonics are considered, the time separation between consecutive attosecond pulses becomes trackable in an energy resolved way. The reshaping of EUV temporal structure near a Cooper minimum in argon is revealed. This wave-front controlled interferometry paves the way for high precision measurement combining both time and spectral resolution with an all-optical approach. It can potentially find significant applications in probing structural information of complex molecular targets.

## Methods

In the experiment the 25fs 800nm laser pulse from the amplifier (Legend Elite HE+ USX) is sent into a hollow-core fiber system filled with neon for spectral broadening. Pulse compression is implemented using double angled chirp mirrors (Ultrafast Innovations PC70). A 7 femtosecond laser pulse with a central wavelength of 760nm is obtained to drive high order harmonics. The static gas cell for harmonic generation is 1 mm long filled with 30 Torr argon gas (or 100 Torr neon gas) . Phase matching of the short trajectory is fulfilled by placing the gas jet 2 mm downstream the laser focus. A small fraction of the total pulse energy is picked off to serve as a perturbing signal field using a broadband beam splitter. It is then recombined with the driver by a compact Mach-Zehnder interferometer. The relative delay between the two pulses can be fine-tuned by a piezoelectric transducer with a delay step of 300 as. The power of the driver and signal can be controlled using diaphragms. Inside the gas cell, the intensity of the driver is about $3\times10^{14}$ W/cm$^2$, and the intensity ratio between the weak perturbing field and the intense driver is maintained below 1%. The high harmonic spectrum is measured with a home built high resolution EUV spectrometer consisting of a 1200 lines/mm flat field grating (Shimadzu 30-002) and a CCD camera (Princeton Instrument PXO-400B). The harmonic spectrum at each delay is accumulated for 5000 laser shots.

We used the Lewenstein model [28] to calculate high order harmonic spectrum. The time dependent dipole moment of an atom in a strong field is given as:

$$D(t) = i\int_0^\infty d\tau \left(\frac{\pi}{\varepsilon+i\tau/2}\right)^{3/2} d^*[P_s(t,\tau) - A(t)]a^*(t) \times \exp(-iS(P_s,t,\tau)) \times E(t-\tau) \cdot d[P_s(t,\tau) - A(t-\tau)] \times a(t-\tau) + c.c.$$

where $d(p)$ is the transition dipole matrix element between the ground state and a continuum state with momentum $p$. $E(t)$ and $A(t)$ represent the electric field and the associated vector potential of the combined driving and signal field, $\varepsilon$ is a positive regularization constant. $P_S$ and $S(P,t,\tau)$ are the canonical momentum and quasi-classical action of the continuum electron at the saddle point, respectively. Ground state depletion is included by introducing the ground-state amplitude $a(t) = \exp\left[-\frac{1}{2}\int_{-\infty}^t w(t')dt'\right]$, where $w(t)$ is the ionization rate calculated with the ADK model [34]. The Fourier transformation of $D(t)$ we denote as the induced dipole $D(\omega)$, and the HHG intensity is proportional to $\omega^4|D(\omega)|^2$.

**Acknowledgements**

This work was supported by the National Natural Science Foundation of China under the grant number 11774111 and the Fundamental Research Funds for the Central Universities under the grant number 2017KFYXJJ142.

**Author contributions:** W.C. and PX. L. conceived and designed the experiment. Z.Y., YL. M. and HY. X. conducted the experiments. Z.Y performed the simulations. All authors participated in the discussions and contributed to the final manuscript.

**Competing interests.** The authors declare no competing interests.

# Wave front controlled attosecond time domain interferometry: supplementary information

Zhen Yang[1], Wei Cao[1], Yunlong Mo[1], Huiyao Xu[1], Kang Mi[1], Pengfei Lan[1], Qingbin Zhang[1] and Peixiang Lu[1,2]

[1]*Wuhan National Laboratory for Optoelectronics and School of Physics, Huazhong University of Science and Technology, Wuhan, 430074, China*

[2]*Hubei Key Laboratory of Optical Information and Pattern Recognition, Wuhan Institute of Technology, Wuhan 430205, China*

We provide important information to support the feasibility of our work in this supplementary material. In section I, the analytical formula of harmonic energy shift under the condition of two slits interfering is derived. In section II, we numerically simulate the high harmonic spectrum using the strong field approximation (SFA) model. The simulated energy shift shows good agreement with the analytical expression and proves the validity of the method. Furthermore, we verify that the derived expression in section 1 is still applicable for multi-slit configuration as well. In section III, we show the measured energy shift is sensitive to the amount of dispersion introduced in the signal field, indicating the feasibility of the current method for reconstructing femtosecond pulse accurately. In section IV, a simple model is utilized to analyze the reshaping of the attosecond pulse train around the Cooper minimum of Ar, which shows evidence of the current interferometry for structure determination. The detailed waveform reconstruction process for elliptically polarized signal pulse is included in section V.

**Section I: Derivation of high harmonic energy shift of double-slit configuration**

When only the driving field is considered and only one short trajectory channel is opened, HHG can be described by SFA where the spectral phase can be written as [1]:

$$\phi(\omega) = -S(P, t_i, t_r) + \omega t_r \qquad (S1)$$

where $P$ is canonical momentum. $t_i$, $t_r$ are the ionization and recombination time of this channel, respectively. ω is the angular frequency, S is the quasi-classical action defined as:

$$S(P, t_i, t_r) = \int_{t_i}^{t_r} \left\{ \frac{[P - A_0(t)]^2}{2} + Ip \right\} dt \qquad (S2)$$

Where $A_0$ is the vector potential of the driving field. $Ip$ is the ionization potential of target gas. The velocity of the electron in field after tunneling can be written as:

$$V(t) = P - A_0(t) \qquad (S3)$$

when a second short trajectory for harmonic generation is introduced, the time interval between two short trajectories is T/2 with T as the optical cycle of the driving field. Considering a CW driving field $E_0(t) = E_0 \sin(\omega_0 t)$, we have:

$$E_0\left(t + \frac{T}{2}\right) = -E_0(t)$$

$$A_0\left(t + \frac{T}{2}\right) = -A_0(t) \qquad (S4)$$

$$P\left(t+\frac{T}{2}\right) = -P(t)$$

Then the quasi-classical action for the second trajectory is:

$$S\left(P, t_r + \frac{T}{2}, t_i + \frac{T}{2}\right) = S(P, t_r, t_i) \tag{S5}$$

When the envelope of the driving field is considered, we have:

$$S\left(P, t_r + \frac{T}{2}, t_i + \frac{T}{2}\right) = S(P, t_r, t_i) + \delta \tag{S6}$$

with $\delta$ accounting for the non-adiabatic condition[2]. Therefore, the phase difference of the two attosecond slits is:

$$\Delta\phi(\omega) = \frac{\omega T}{2} + \pi - \delta \tag{S7}$$

A phase $\pi$ is introduced to consider the sign flip of the electric field that drives the two consecutive attosecond pulses(slits). When a perturbing signal field is introduced, the phase shift of the first slit has been presented in [3] as:

$$\sigma_1(\tau') = -\frac{1}{72} E_0 \omega_0 t_d^4 E_s(\tau' + \xi) \tag{S8}$$

Where $t_d$ is the excursion time of the electron, $\xi = 2t_r/5$ is a constant time shift [3]. $\tau'$ is the delay between the driving and the signal field. The similar derivation can be applied to the second slit:

$$\sigma_2(\tau') = \frac{1}{72} \alpha E_0 \omega_0 t_d^4 E_s(\tau' + \xi + \Delta) \tag{S9}$$

Here, we set $\tau = \tau' + \xi$ as the relative delay for convenience. $\Delta$ is the time interval between two attosecond slits. The phase shifts of two slits are opposite as the vectors of driving field are opposite between the two slits. The coefficient $\alpha$ is introduced consider the intensity variation of the driving field within an optical cycle. In consequence, the coefficient $\alpha$ represents the intensity ratio of driving field at two consecutive half cycles. Finally, the total phase difference of the two slits can be written as:

$$\begin{aligned}\Delta\Phi(\omega, \tau) &= \sigma_2(\tau') - \sigma_1(\tau') + \Delta\phi(\omega) \\ &= \frac{E_0 \omega_0 t_d^4}{72} \left(\alpha E_s(\tau + \Delta) + E_s(\tau)\right) + \frac{\omega T}{2} + \pi - \delta\end{aligned} \tag{S10}$$

When $\Delta\Phi(\omega, \tau) = 2m\pi$ with $m$ as an integer, two slits interfere constructively. We have:

$$\frac{E_0 \omega_0 t_d^4}{72} \left(\alpha E_s(\tau + \Delta) + E_s(\tau)\right) + \frac{\omega T}{2} + \pi - \delta = 2m\pi \tag{S11}$$

Hence, the high order harmonics will peak at:

$$\omega = (2m-1)\omega_0 + \frac{2\delta}{T} - \frac{E_0 \omega_0 t_d^4}{36T} \left(\alpha E_s(\tau + \Delta) + E_s(\tau)\right) \tag{S12}$$

both $m$ and $\delta$ are constants and independent with the signal field, therefore the expression of energy shift caused by the signal filed is:

$$\sigma(\tau) = -\frac{E_0 \omega_0 t_d^4}{36T} \left(\alpha E_s(\tau + \Delta) + E_s(\tau)\right) \tag{S13}$$

**Section II: The reconstruction of the signal field and the effect of multi-slit configuration**

We performed the numerical simulation of HHG using SFA to validate the expression we derived in section I. An intense ($2 \times 10^{14}$ W/cm²) three-cycle 800nm driving field interacts with target Ar for high harmonic generation (HHG).

A weak ($1 \times 10^{12}$ W/cm$^2$) two-cycle 800nm signal field is introduced to perturb HHG. Figure 1(a) shows the two dimensional spectrogram of the high order harmonic radiation by scanning the delay between the two fields. In figure 1(b), we compared the energy shift simulated from SFA with that from Eq. (S13). They show good agreement while non-adiabatic condition is taken into account, while the small discrepancy is caused by the high order term from Eq. (S2) ignored in the derivation[3]. Figure 1(c) shows the Fourier transform of delay dependent energy shift simulated by SFA near 42 eV. A dip near $\omega_d = 1.55 eV$ is due to destructive interference of the two delayed signals depicted in equation (S13) and is directly related to the time interval between the two attosecond slits $\Delta = \frac{\pi}{\omega_d}$. With $\Delta$ known, it is straightforward to extract the complete information of the electric field E$_s$: $E_s(\omega) \propto \frac{\sigma(\omega)}{(1+\alpha e^{i\Delta\omega})}$, where $\alpha$ is adjusted to keep the spectrum of the reconstructed field similar with that of the original signal field and is in general close to 1. An inverse Fourier transform gives the time domain electric field of the signal pulse as is shown in figure 1(d). It shows that the reconstructed and original signal field is in excellent agreement, indicating the feasibility of our method. For expanding the application, we increase the number of slits by adjusting the window function, which is applied for selecting the number of short trajectories in SFA. We have compared the energy shifts simulated by SFA where multi-slit condition is considered and they appear to be almost identical (figure. 2). It indicates Eq. (S13) is still feasible for multiple slits. This can be easily achieved when an ultrashort driving pulse is used for high harmonic generation.

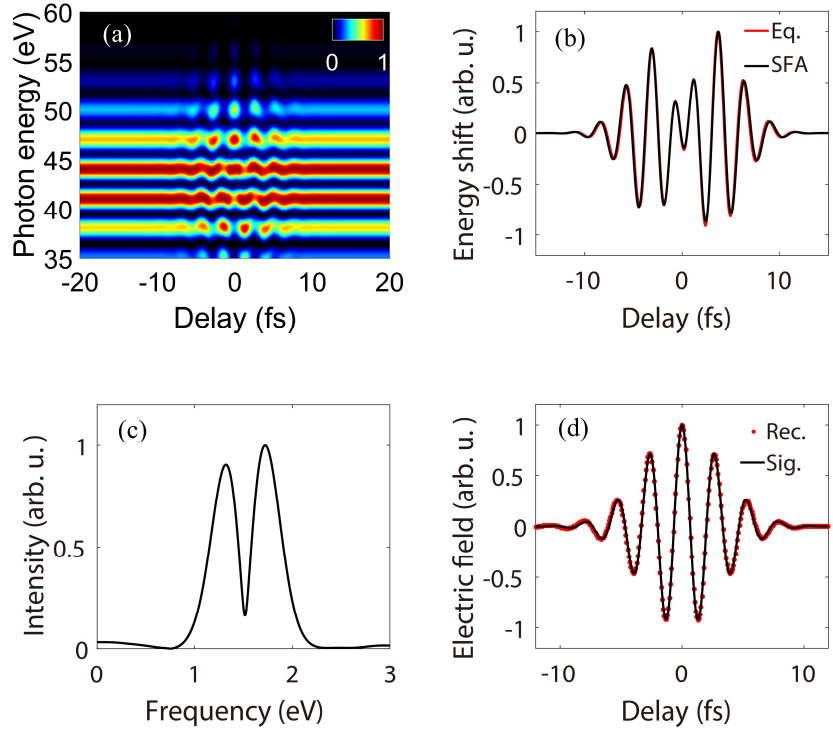

**Figure 1** (a) Two dimensional spectrogram of the high order harmonic radiation simulated by SFA. (b) The normalized results of energy shift extracted from (a) near 42 eV (black curve) and calculated using Eq. (S13)(red curve). (c) The Fourier transform of delay dependent energy shift in (b). (d) The reconstructed field (red dot curve) and the original signal field (black curve).

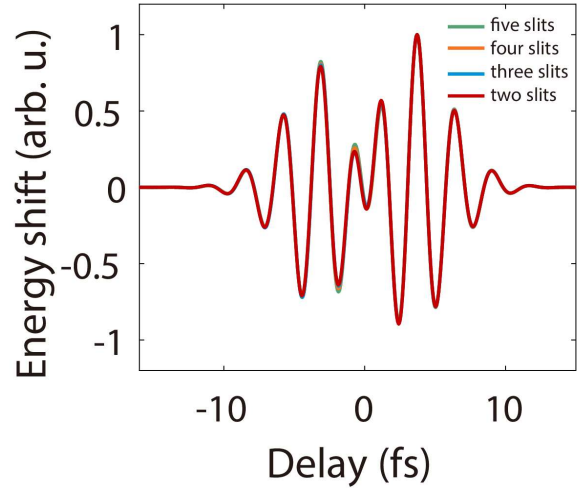

**Figure 2** The energy shift of the 27$^{th}$ harmonic (near 42 eV) for different number of attosecond slits. The parameters are the same as figure 1.

**Section III: Experimental results for chirped signal pulses**

In the experiment, we have also altered the waveform of the signal field by introducing dispersion materials to test its generality. The delay dependent centroid of the argon harmonic near 38 eV for different thickness of fused silica inserted in the signal arm is inspected. The time-frequency analysis of the delay dependent energy shift is shown in figure 3. The position of the dip around 1.55 eV is again due to the destructive interference as mentioned in section II. As the fused silica is getting thicker, a progressive tilt of the spectrogram is observed, indicating that the signal field is gaining a positive group delay dispersion. It implies that our scheme is sensitive enough for capturing the carrier of the signal pulse.

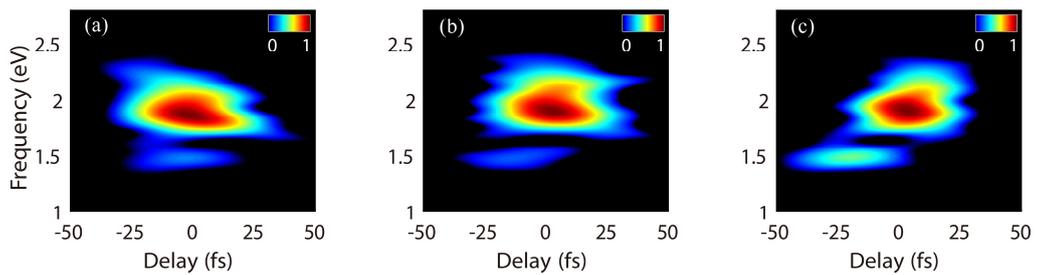

**Figure 3** The time-frequency analysis of measured energy shift of harmonic near 38 eV after inserting 0mm (a), 0.3mm (b) and 0.5mm (c) of fused silica in the signal arm.

**Section IV: The reconstruction of elliptically polarized signal pulse by attosecond few-slit interferometry**

In this section, we introduce the detailed process of measuring vectorial optical field. In general, a vectorial laser field can be divided into two perpendicularly polarized components. By measuring the two components independently, the total vectroial field can be reconstructed. In the present of a vectorial signal field, the total phase difference of the two slits can be written as:

$$\Delta\phi(\omega) = S\left(\vec{P}, t_i + \frac{T}{2}, t_r + \frac{T}{2}, \tau\right) - S(\vec{P}, t_i, t_r, \tau) + \frac{\omega T}{2} + \delta$$

$$\approx \int_{t_i}^{t_r}\left\{\left[\vec{P} - \vec{A_d}\left(t + \frac{T}{2}\right)\right]\vec{A_s}\left(t + \frac{T}{2} + \tau\right)\right\}dt -$$

$$\int_{t_i}^{t_r}\{[\vec{P} - \vec{A_d}(t)]\vec{A_s}(t + \tau)\}dt + \frac{\omega T}{2} + \pi + \delta \qquad S(14)$$

Where $\vec{A_d}$, $\vec{A_s}$ and $\vec{P}$ is the vector potential of driving field, the vector potential of signal field and canonical momentum almost equal to $\vec{P_0}$ ($\vec{P_0}$ is the canonical momentum parallel to $\vec{A_d}$ when only driving field exists) when the signal field can be regarded as a perturbation, respectively. When the driving field is linearly polarized, only the component of signal field polarized parallel to the driving field contribute to the perturbation of HHG [4, 5]. Thus, the component parallel to the driver polarization can be retrieved by the reconstruction scheme mentioned in section II. Thus, the two components of the signal pulse can be measured by rotating the polarization of the signal field or driving field by 90°. We choose the latter in our simulation for convenience. Fig. 4 shows the results simulated by SFA with the same condition mentioned in section II except that the signal field is circularly polarized instead. Figure 4(a) and (c) shows the two dimensional spectrograms of HHG perturbed by circularly polarized signal field when the polarization direction of driving field is along the horizontal (x) and vertical(y) direction, respectively. Fig. 4(b) and (d) shows the reconstructed horizontal and vertical components of signal field from the energy shift extracted from (a) and (c) near 42 eV. It should be noticed that the reconstructed signal field has been normalized to the x component. Figure 4(d) indicates that the reconstructed circularly polarized signal field is consistent with the original one.

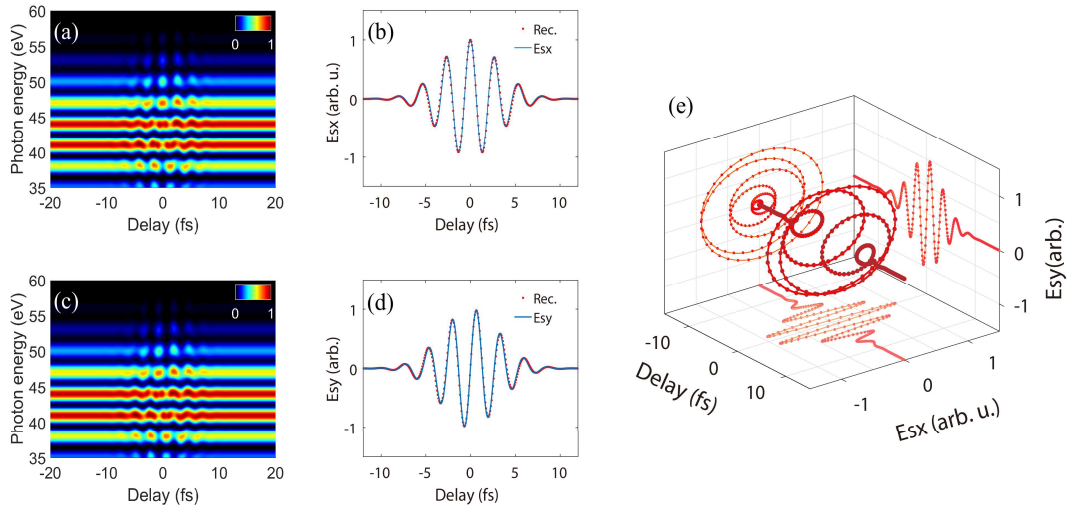

**Figure 4** (a) Calculated high order harmonic spectrum perturbed by a circularly polarized signal field with the driver polarized along x direction. (b) The reconstructed (red dots) and the original (blue solid line) x component of the signal field using harmonic near 42 eV. (c) Calculated high order harmonic spectrum perturbed by a circularly polarized signal field with the driver polarized along y direction. (d) The reconstructed (red dots) and the original (blue solid line) y component of the signal field using harmonic near 42 eV. (e) The reconstructed (orange dots) and the original (red solid line) waveform of the circularly polarized signal field.

In the experiment we rotate the signal field by 90° while keeping the polarization of the driver unchanged and performed two independent measurements for reconstructing the waveform of elliptically polarized signal pulses. A remarkable issue is that we should keep the delay axis the same in the two independent experiments. To achieve this purpose, we set up an interferometer locking system. Our apparatus is a Mach-Zehnder interferometer configuration (figure 5). A weak 532 nm continuous wave (CW) green laser is introduced to co-propagate with the infrared field. The interference of the green laser is serving as a feedback for locking [6] the interferometer as well as providing a reliable reference to determine a common delay axis for the two measurements. The Mach-Zehnder interferometer after adding the feedback system can be locked to within 15 as RMS (the inset in figure 5).

The strong driving pulse is horizontally polarized (we define horizontal direction as x direction) in our experiment. A zero order quarter-wave plate and a zero order half-wave plate (750nm) are inserted in the signal arm for polarization control. The optical axis of both wave plates are horizontal such that both the diving and the signal pulse are horizontally polarized. A circularly or elliptically polarized signal can be generated by rotating the quarter-wave quartz by a certain angle. After that, by rotating the half-wave quartz, the two orthogonal polarization components of the signal field can be selected to match the polarization of the driver for diagnosis. It should be noticed that rotating the half-wave plate will induce a drift in the interference pattern of green laser as well because of the refractive index change of the wave plates for green light. We have taken this effect into account in order to determine an accurate common delay axis for the two independent measurement. The corrected two dimensional spectrogram of HHG perturbed by the circularly (a1, a2) and elliptically polarized signal fields (b1, b2) are shown in figure. 6.

It is shown that the energy shift induced by vertical component is comparable to that of horizontal component for circularly polarized signal pulse. In contract, the energy shift induced by vertical component is weaker than that of the horizontal component for elliptically polarization signal pulse (see figure 5 (c) in main paper). Thus our method is sensitive to the phase and amplitude ratio of the two components, and can measure waveform of the signal field with great accuracy. The reconstructed elliptically polarized signal field are shown in figure 4 of the main paper.

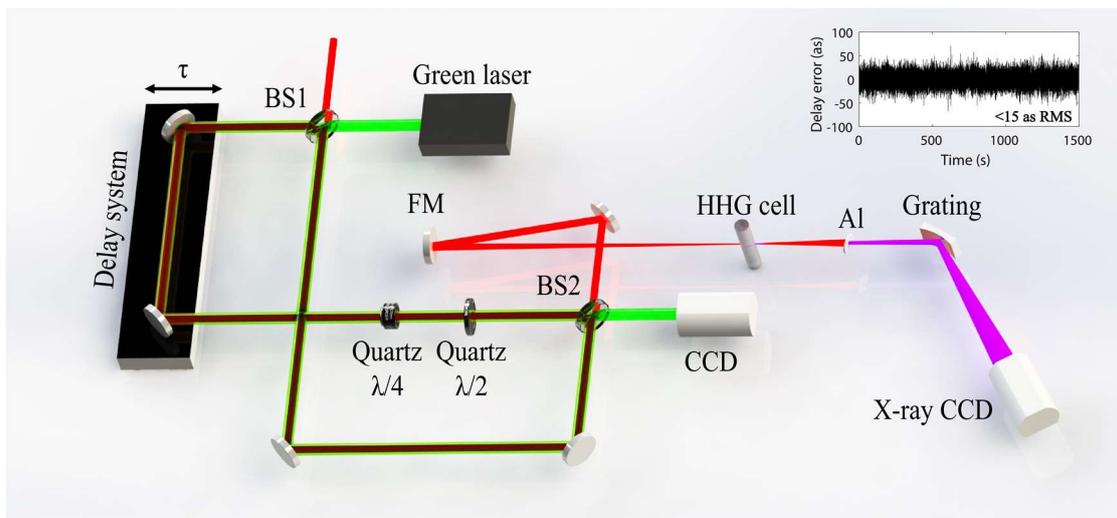

**Figure 5** Experimental set-up. BS: beam splitter, FM: focusing mirror. Al: aluminum filter for blocking the remained infrared light. Inset: The relative delay jittering between the two arms is within 15 as in our Mach-Zehnder interferometer.

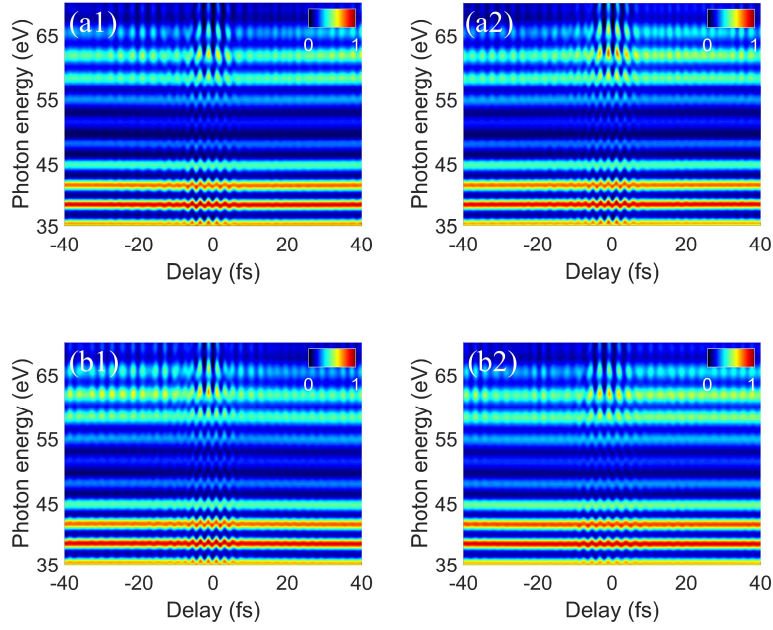

**Figure 6** The measured high order harmonic spectra perturbed by the horizontal(a1) and vertical(a2) component of a circularly polarized signal field, and by the horizontal(b1) and vertical(b2) component of an elliptically polarized signal field. For circularly polarized light, quarter wave plate is rotated by 45°. For elliptically polarized light, quarter wave plate is rotated by 30°(ellipticity equals to 0.58 in theory).

**Section V: Model calculation of the reshaping of attosecond pulse train around the Cooper minimum of Ar**

In general, the time interval of two consecutive EUV pulses is close to T/2, considering a smooth phase distribution over the harmonic spectrum, i.e. the recombination dipole matrix element has a smooth behavior. However, when a phase jump appears in the recombination dipole matrix element, as is shown around the Cooper minimum(CM) of argon, it will alter the shape of the emitted attosecond pulses [7]. Thanks to the energy resolution provided by the current interferometric technique, the separation of the two slits can be inspected in an energy resolved manner around the Cooper minimum. Any subtle variation in the time interval of the attosecond pulse train can be sensitively captured by interrogating the destructive interference minimum of the power spectrum of the harmonic energy shift trace, thus offering a way to study the structural features of the atomic or molecular targets. To verify this, a simple model calculation is presented here. In the simulation, both the driving field ($3\times10^{14}$ W/cm²) and the signal field($2\times10^{12}$ W/cm²) are 800nm with a pulse duration of two optical cycles . The atomic structure is included in the calculation by using different dipole matrix elements in the SFA model. To mimic the HHG of neon, we used the hydrogenlike transition dipole: $d(p)_H = \left(\frac{2^{7/2}\alpha^{5/4}}{\pi}\right)\frac{p}{(p^2+\alpha)^3}$ with p standing for the canonical momentum. In case of argon where a Cooper minimum exists around 51 eV, we used the transition dipole in the form similar to reference[7]: $d(\omega) = d(\omega)_H \left[1 + \frac{\omega_C - \omega}{\Delta\omega} e^{i\xi(\omega)}\right]$ with $\omega_C = 45.6eV$, $\Delta\omega = 6.3eV$, $\xi(\omega)=1.4$ rad. $\omega$ represents the photon energy and is related to p as: $\frac{h\omega}{2\pi} = \frac{p^2}{2} + I_p$ with $I_p$ the ionization potential of the generating atom. This transition dipole has a phase jump of 2.2 rad across a range of 20 eV centered at 51 eV and creates a minimum in the harmonic spectrum yield around 51 eV that resembles the main features of argon HHG. We applied a window function to select the short trajectories in the final dipole moment. Harmonic spectra for atom without abnormal atomic structure is shown in figure 7(b), the Fourier analysis of the energy shift of individual harmonic (figure 7(d)) gives a common minimum position around 1.55 eV that is independent on the harmonic order. Harmonic spectra for atoms with an abnormal atomic structure is shown in figure 7(a), and a minimum around 51 eV in the harmonic yield is clearly observed and indicates the CM. Figure 7(c) shows the power spectrum of the energy shift of individual harmonics. The minimum position in the power spectrum shifts towards the higher frequency components around the CM, which is consistent with the experimental results as shown in the main text. Therefore, this energy resolving capability of the current interferometry allows the probing of abnormal structure of atoms.

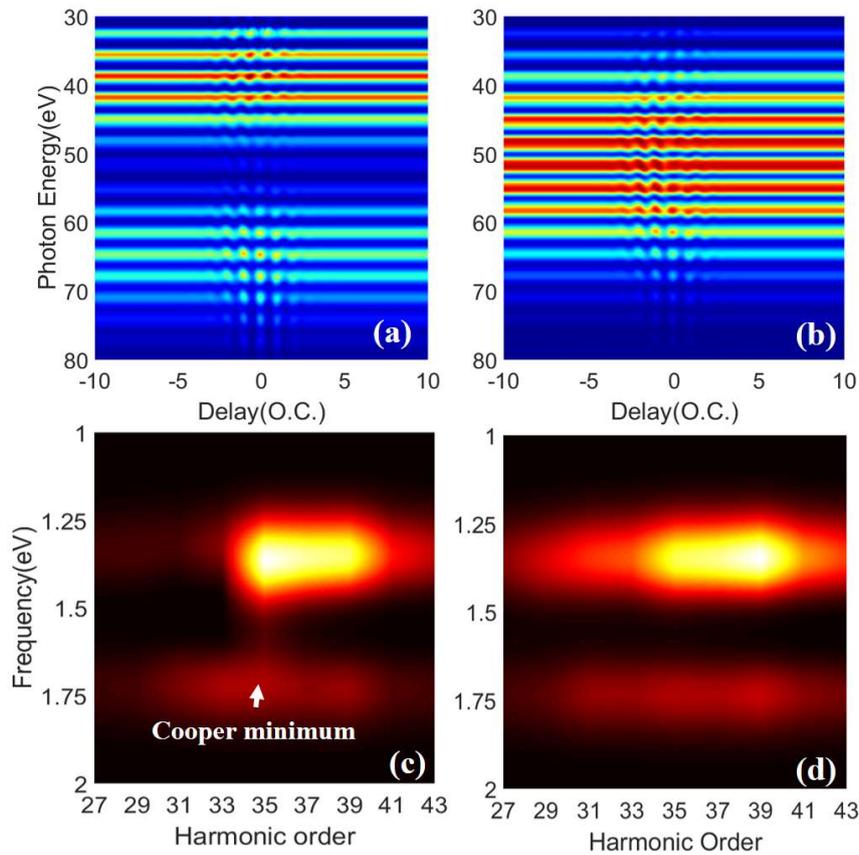

**Figure 7** Model calculation of high order harmonic spectra perturbed by a signal pulse for two different bound-free transition dipole structures. In (a), the bound-free transition dipole undergoes a phase jump near 51 eV (Cooper minimum). (c) The Fourier analysis of the delay dependent energy shift for harmonics in (a) from $27^{th}$ to $43^{rd}$ order. In (b), the bound-free transition dipole has a flat phase. (d) The Fourier analysis of the delay dependent energy shift for harmonics in (b) from $27^{th}$ to $43^{rd}$ order.